\documentclass[pss]{wiley2sp} 
\usepackage{amsmath}
\usepackage{bm}             

\begin{document}

\title{STABILIZATION OF THE PEROVSKITE PHASE IN THE Y-Bi-O SYSTEM BY USING A BaBiO$_{3}$ BUFFER LAYER}

\author{%
	Rosa Luca Bouwmeester\textsuperscript{\Ast,\textsf{\bfseries 1}},
 	Kit de Hond\textsuperscript{\textsf{\bfseries 1}},
  	Nicolas Gauquelin\textsuperscript{\textsf{\bfseries 2}},
  	Jo Verbeeck\textsuperscript{\textsf{\bfseries 2}},
  	Gertjan Koster\textsuperscript{\textsf{\bfseries 1}},
  	Alexander Brinkman\textsuperscript{\textsf{\bfseries 1}}
  	}

\mail{e-mail
  \textsf{r.l.bouwmeester@utwente.nl}}

\institute{%
  \textsuperscript{1}\,MESA$^+$ Institute for Nanotechnology, University of Twente, The Netherlands\\
  \textsuperscript{2}\,Electron Microscopy for Materials Research (EMAT), University of Antwerp, Belgium}


\abstract{\bf%
A topological insulating phase has theoretically been predicted for the thermodynamically unstable perovskite phase of YBiO$_{3}$. Here, it is shown that the crystal structure of the Y-Bi-O system can be controlled by using a BaBiO$_{3}$ buffer layer. The BaBiO$_{3}$ film overcomes the large lattice mismatch of 12\% with the SrTiO$_{3}$ substrate by forming a rocksalt structure in between the two perovskite structures. Depositing an YBiO$_{3}$ film directly on a SrTiO$_{3}$ substrate gives a fluorite structure. However, when the Y-Bi-O system is deposited on top of the buffer layer with the correct crystal phase and comparable lattice constant, a single oriented perovskite structure with the expected lattice constants is observed. }

\maketitle   

\section*{Introduction} \text{ }
Topological materials are at the focus of a relatively young field in material science, aiming to find both new topological materials as well as discovering novel applications for this new class of matter. The most prominent phenomena that emerge in topological materials include a chiral spin structure and topologically protected surface states [1, 2]. The spin-momentum locking makes topological matter interesting for spintronic applications [1, 3, 4]
and possibly quantum computation [5]. Therefore, global research is conducted towards further developing known topological materials and trying to find new compounds that have a non-trivial topology in their band structure.

So far, only a handful of materials have been identified as topological insulators~(TIs). Within a year after the theoretical prediction of Bernevig \textit{et al.} [1], K\"onig \textit{et al.} [3] observed the quantum spin Hall state in CdTe/HgTe/CdTe quantum wells - the first system known with a non-trivial band structure. Bi$_{1-x}$Sb$_{x}$ was the first three-dimensional TI that was experimentally observed [6], followed by Bi$_{2}$Se$_{3}$ [7] and Bi$_{2}$Te$_{3}$ [4]. The relatively heavy bismuth gives rise to strong spin-orbit coupling (SOC) effects, which in turn results in a band inversion at an odd number of time reversal invariant momenta (TRIMs) and a non-trivial topological band structure [8].

Almost all of the experimentally confirmed TIs are either metallic in nature or formed by quantum wells. Therefore, it is useful to investigate whether the field of TIs can be expanded into other classes of materials, such as complex oxides. The large tunability of the growth processes of oxides and the wide range of possible materials opens up a whole new realm of unexplored possible TIs. Several oxides have theoretically been predicted to be TIs, most of them containing heavy elements such as bismuth, but experimental verification has remained unsuccessful thus far.

In 2013, Jin \textit{et al.}~[9] predicted YBiO$_{3}$ (YBO) to be a topological insulator, implying a non-trivial band topology. The electronic band structure was theoretically calculated while the SOC strength was increased. A gapless Dirac cone was found in the band structure of YBO, which indicates a topological phase transition. A lattice constant of 5.428~\r{A} was used in all the calculations and a simple cubic pervoskite structure was adopted. This lattice constant was experimentally obtained by Li \textit{et al.} [10] with X-ray diffraction measurements. However, the lattice constant was assigned to a cubic structure, but the related crystal phase was never specified.

By theoretically reproducing the result of Jin \textit{et al.} [9], Trimarchi \textit{et al.} [11] found that the total energy of the system drops when the lattice constant is allowed to be relaxed. The new equilibrium value for the perovskite lattice constant is \textit{a}=4.405~\r{A}. When the yttrium and bismuth positions were no longer set as a constraint, the total energy was lowered even further by interchanging the occupation of the two cation sites. The BiYO$_{3}$ (BYO) configuration has a lattice constant predicted to be 4.349~\r{A}. This, however, is no longer a TI, but a trivial insulator.

\begin{figure*}[htb]%
\sidecaption
\includegraphics*[width=0.7\textwidth]{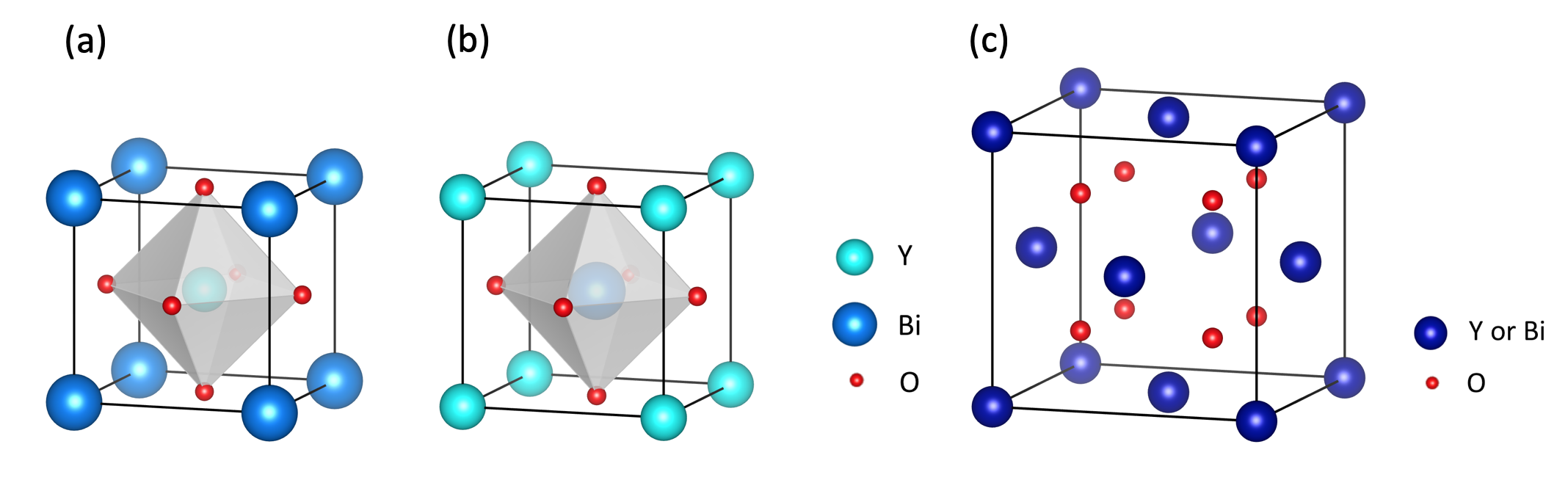}
\caption{(a), (b) and (c) show the perovskite YBiO$_{3}$, BiYO$_{3}$ and the fluorite YBiO$_{4}$, respectively. The red dots represent the oxygen atoms, the light blue the yttrium atoms, the medium blue dot the bismuth atoms and the dark blue dots either the yttrium or bismuth atoms in the fluorite phase. For a better visualisation of the structures, the atoms are not depicted in their realistic radii.}
    \label{fig:Scenarios}
\end{figure*}

Based on X-ray powder diffraction experiments, Zhang \textit{et al.}~[12] demonstrated the Y-Bi-O system to be stable in the high temperature metastable cubic phase as YBO in the fluorite structure, which implies that 25\% of the oxygen positions are unoccupied -~a fluorite structure with two cations has a general formula ABO$_{4}$ (see figure \ref{fig:Scenarios}(c)). The experimentally observed lattice constant was $\sim$5.4~\r{A}. This results is not in agreement with the perovskite phase adopted in the calculations performed by Jin \textit{et al.} [9]. With UV-vis absorption spectra a direct band gap of 5.88~eV is determined, which excludes the possibility of a topological insulating phase in the fluorite YBO.

By density functional theory (DFT) calculations Zhang \textit{et al.}~[12] found YBiO$_{3}$ and BiYO$_{3}$ as possible outcomes (see figures \ref{fig:Scenarios}(a) and (b), respectively) when a simple perovskite structure was given as a constraint. The lattice constants are 4.373~\r{A} and 4.283~\r{A} for the perovskite YBO and BYO, respectively, close to the values found by Trimarchi \textit{et al.} [11]. The band structure of BYO shows an insulating structure without a Dirac point, so it cannot be a TI. 

In the calculations of the band structure of perovskite YBO, Zhang \textit{et al.} [12] also included a spin-orbit coupling strength, just as Jin \textit{et al.} [9] did in their calculations. A ground state with a metallic character was found, since a band crosses the Fermi level. However, in both theoretical calculations, the SOC strength is not quantified. If SOC is strong enough, the perovskite YBO can still be a TI. In table \ref{tab:ybo}, the four scenarios are summarized.

\begin{table}
  \caption{The four possible scenarios of the Y-Bi-O system. The crystal structure with the corresponding lattice constant (\textit{a}) and the possibility if the system is a topological insulator (TI) or not.}
  \begin{tabular}[htbp]{@{}lllll@{}}
    \hline
	Compound & TI? & Crystal structure & \textit{a} & Ref. \\
    \hline
YBiO$_{3}$ & No & Perovskite & 5.4238 \r{A} & 9   \\
    YBiO$_{3}$ & No
    & Fluorite & $\sim$ 5.4 \r{A} & 12  \\
    YBiO$_{3}$ & Yes
    & Perovskite & $\sim$ 4.4 \r{A} & 11, 12 \\
     BiYO$_{3}$ & No
    & Perovskite & $\sim$ 4.3 \r{A} & 11, 12 \\
    \hline
  \end{tabular}
\label{tab:ybo}
\end{table}

Here, we show control over the crystal phase of YBO for thin film depositions. To realize the energetically unfavourable perovskite phase, the use of a buffer layer is investigated. In literature buffer layers are often used to control the in-plane strain [13, 14], quality [15, 16] and crystal orientation [17] of the film deposited on top. BaBiO$_{3}$ (BBO) is suggested as suitable buffer layer material, since it has a lattice constant (\textit{a}=4.35 \r{A} [18]) comparable to the predicted perovskite YBO and it is stable in the perovskite phase. Furthermore, it has an insulating character, required to be able to detect possible conducting surface states at the YBO/BBO interface.

\section*{Experimental} \text{ }
BBO and YBO thin films were deposited on a TiO$_{2}$ terminated SrTiO$_{3}$(001) (STO) substrate (CrysTec GmbH) with pulsed laser deposition (PLD) and characterized by \textit{in-situ} reflection high-energy electron diffraction (RHEED) and X-ray photoelectron spectroscopy (XPS). Additional \textit{ex-situ} characterization was performed with an X-ray diffraction (XRD) setup and high resolution scanning transmission electron microscopy (HRSTEM). The substrates were prepared by a wet etching step in a buffered hydrogen fluoride (BHF) solution, followed by a 1.5~hour annealing step at 950$^{\circ}$C [19].

Stoichiometric targets of Ba-Bi-O (house-made, purity 99.99\%) and Y-Bi-O (SurfaceNet GmbH, purity 99.99\%) were used during the pulsed laser deposition with a KrF laser at a fluency of 1.9 J/cm$^{2}$, while the substrate temperature was 500$^{\circ}$C and the O$_{2}$ background pressure was set to 10$^{-2}$ mbar. The distance between the target and substrate was kept constant at 50~mm. A repetition rate of 1~Hz and a total of 900 pulses was used for the BBO buffer layers. The YBO films were grown using interval deposition, which implies that 30 pulses are fired as a bunch at a frequency of 50 Hz followed by a 30 seconds wait interval allowing the deposited atoms to rearrange and form a film. Layer-by-layer growth can be controlled during a PLD with interval deposition as described by Koster \textit{et al.} [20].

These PLD parameter settings differ from previous research. Previously, Orsel \textit{et al.}~[21] performed laser-induced fluorescence (LIF) spectroscopy during the YBO deposition and mapped the spatiotemporal distribution of Y, YO, Bi and BiO constituents while the composition of the background pressure was changed from a pure argon gas to a pure oxygen background gas, keeping the total pressure constant at 0.100 mbar. By XPS and XRD measurements, it was shown that non-epitaxial YBO appears when the background pressure consist for 80\% or 100\% out of O$_{2}$. Below 20\% O$_{2}$ in the total background pressure composition, no Bi was found in the films. This originates from reactions between the target and background gas and not from chemical interactions of the plasma plume with Ar and O$_{2}$ particles. By now combining a decreased total background pressure, as compared to Orsel \textit{et al.} [21], and the use of interval deposition we show that it is possible to deposit stoichiometric YBO films in a pure O$_{2}$ environment.

\section*{Results} \text{ }
\begin{figure*}[htb]%
\sidecaption
\includegraphics*[width=0.55\textwidth]{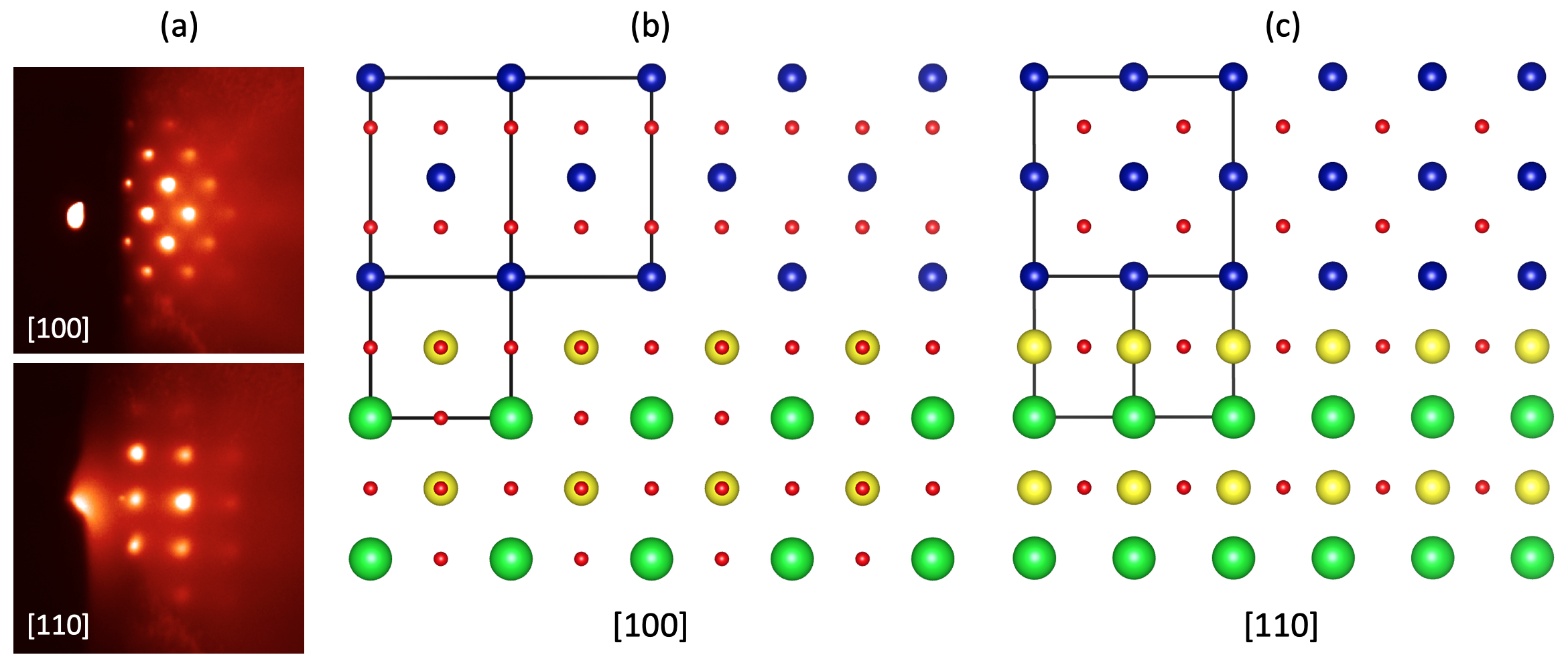}
\caption{(a) The RHEED patterns of a fluorite YBO film on a STO substrate, captured along the [100] and [110] of the substrate for top and bottom, respectively. (b) and (c) visualize the fluorite YBO structure grown on a perovskite STO substrate, both depicted from the side. In (b) the substrate is shown along the [100] direction and the YBO in the [110] direction. In (c) the structure is rotated 45$^{\circ}$ with respect to (b), so that the STO is in the [110] and the YBO in the [100]. The green, yellow, blue and red dots represent the strontium, titanium, yttrium/bismuth and oxygen atoms, respectively. The atoms are not depicted in their realistic radii.}
\label{fig:YBOFluorite}
\end{figure*}
If an YBO film is deposited directly on a STO substrate, it stabilizes in the fluorite phase, as shown by the RHEED patterns in figure \ref{fig:YBOFluorite}(a). The top and bottom RHEED patterns are taken along the [001] and [011] directions of the substrate, respectively. The fluorite structure is rotated 45$^{\circ}$~in-plane with respect to the substrate. In figure \ref{fig:YBOFluorite}(b) and~(c) the substrate with the fluorite YBO film on top are schematically shown along the [001] and [011] directions of the substrate, respectively. The RHEED pattern and schematic structure along the [001] direction of the substrate, both show an elongated YBO structure. When the YBO film is rotated by 45$^{\circ}$ within the RHEED bundle - so that the film is show along the [011] direction of the substrate, a square-like pattern of the YBO is observed as in agreement with the schematic structure shown in (c).

The 45$^{\circ}$ rotation of the YBO film is explained by the matching oxygen planes of the perovskite STO phase and the fluorite YBO phase, as shown by O$'$Sullivan~\textit{et al.} [22] who also grows a fluorite phase on a perovskite structure. The distances between the oxygen atoms in the perovskite STO substrate and fluorite YBO structures are similar, the difference is solely $\sim$2\%. Therefore, only a small strain is needed to epitaxially grow these structures on top of each other when the fluorite YBO is rotated 45$^{\circ}$ with respect to the STO substrate. 

\begin{figure*}[htb]%
\sidecaption
\includegraphics*[width=0.73\textwidth]{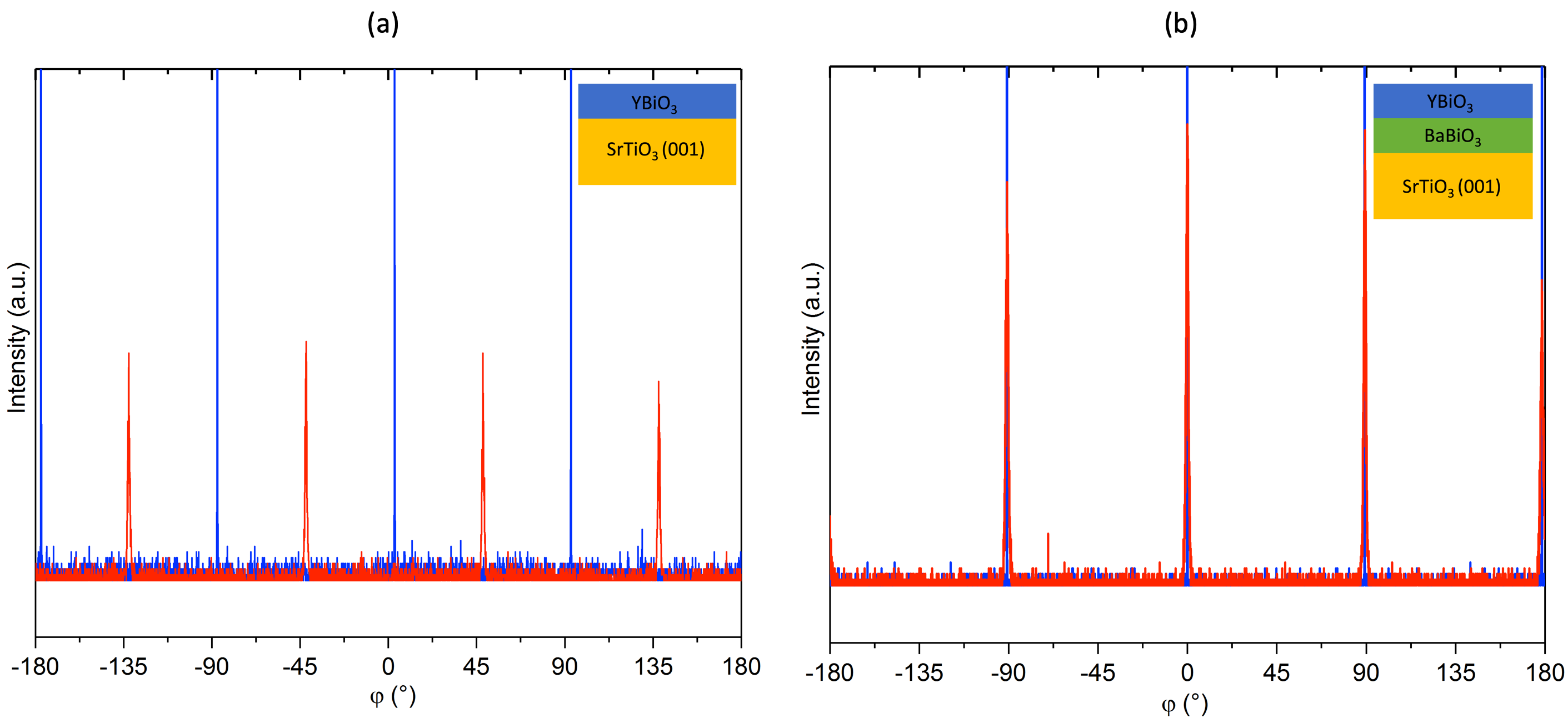}
\caption{Azimuthal scans of (a) an YBO film deposited directly on the STO substrate and (b) with a BBO buffer layer in between the YBO film and the substrate, as shown by the schematics in the top right corners. The blue peaks indicate the STO $<$101$>$ diffraction peaks. In (a) the red peaks show the $<$202$>$ diffraction peaks of the fluorite YBO and in (b) these indicate the $<$101$>$ diffraction peaks of both the perovskite YBO and BBO films.}
    \label{fig:YBOPhi}
\end{figure*}

Azimuthal scans are performed and shown in figure~\ref{fig:YBOPhi}(a), where the STO $<$101$>$ diffraction peaks (in blue) and the fluorite YBO $<$202$>$ diffraction peaks (in red) are observed. The scan confirms the 45$^{\circ}$ rotation of the fluorite YBO on the STO substrate and shows the four-fold symmetry of both structures. The element ratio, determined with an \textit{in-situ} XPS, confirms the oxygen deficient fluorite phase. The Y:Bi:O element ratio is 22:19:59$\pm$3\%, respectively. The XPS data is given in the supporting information, figure S1(a).

\begin{figure*}[htb]%
\includegraphics*[width=\textwidth]{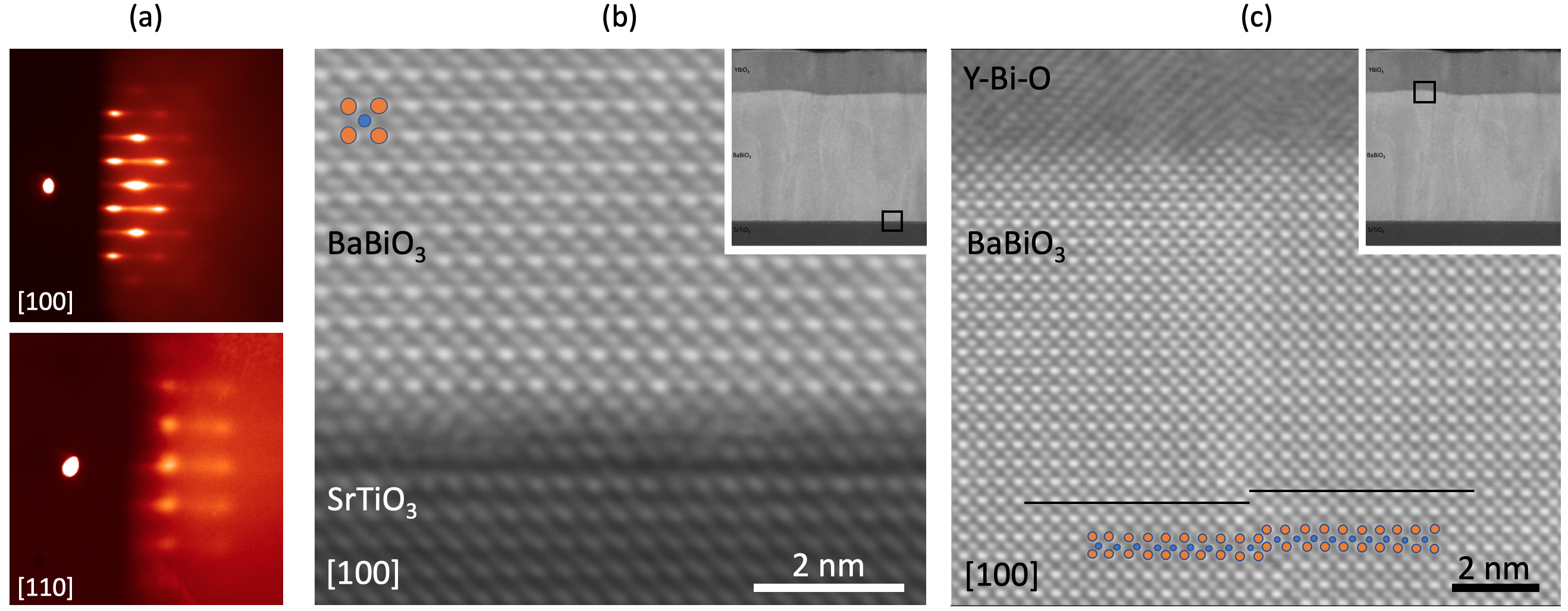}
\caption{(a) RHEED patterns of a BBO buffer layer on a STO substrate taken along the [100] and [110] of the substrate for top and bottom, respectively. (b) HAADF-STEM image of the interface between a STO substrate and the BBO film. (c) HAADF-STEM image of the top of the BBO film and its interface with the YBO top layer. In both (b) and (c) the overview image of the full specimen, with a small box showing the location of the region presented in the main panel, is shown in the top right corner. In both images Bi atoms are represented by the orange dots and Ba atoms by the blue dots. In (c) the two black lines follow the bismuth atoms, indicating the anti-phase boundary where the unit cell shifts by half a unit cell.}
    \label{fig:BBOTEM}
\end{figure*}

\begin{figure*}[htb]%
\sidecaption
\includegraphics*[width=0.6\textwidth]{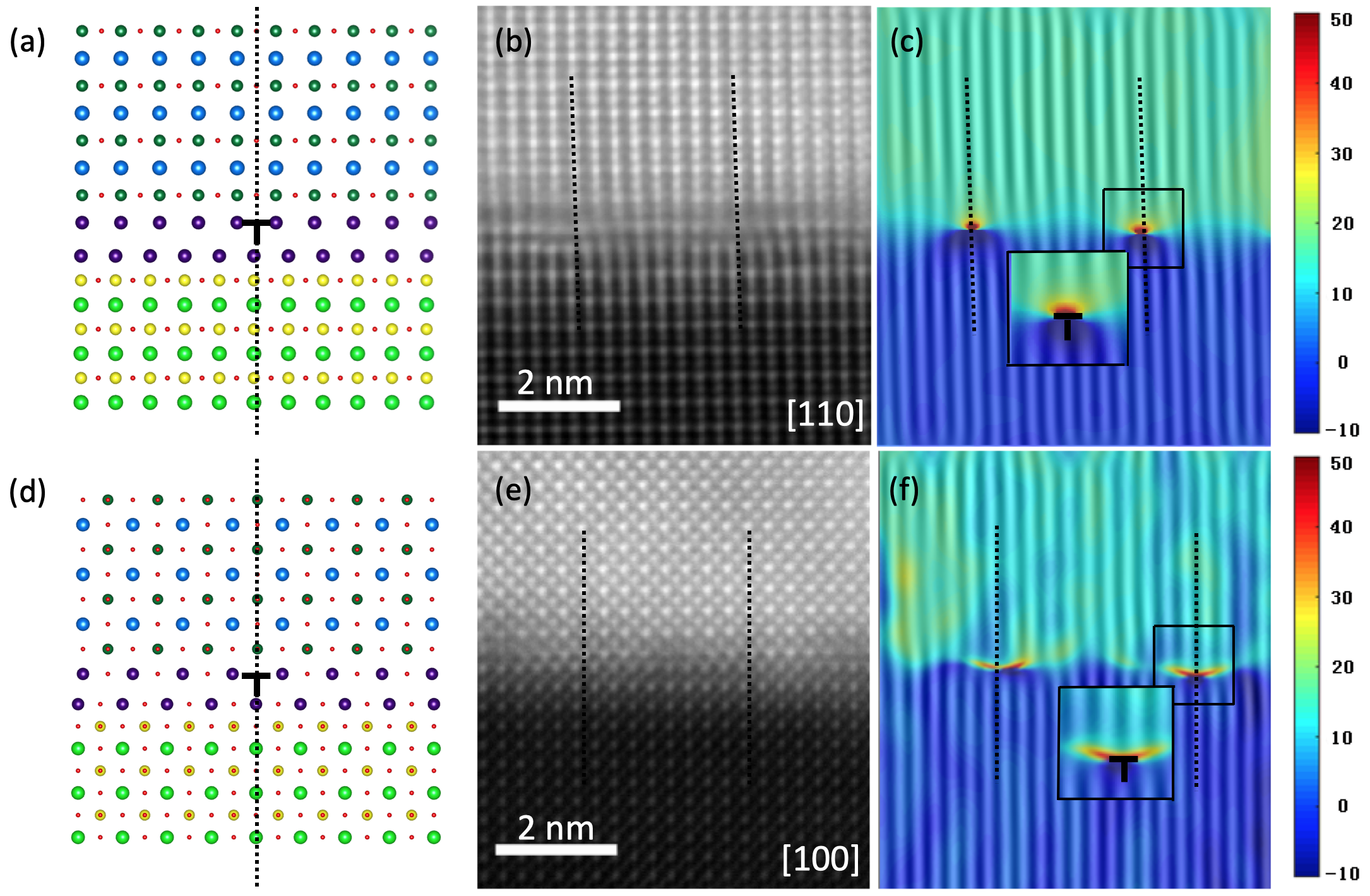}
\caption{Strain relaxation at the BBO/STO interface. (a) and (d) present schematic models of the STO substrate, the two AO layers and the BBO film along the [110] and [100] directions, respectively. The green, yellow, blue, dark green and red dots represent the strontium, titanium, barium, bismuth and oxygen atoms, respectively. The purple dots indicate the cation sites in the two AO layers. The atoms are not depicted in their realistic radii. In (b) and (e) the HAADF STEM images of the BBO/STO interface are shown along the [110] and [100] zone axis, respectively. (c) and (f) show the GPA analysis images overlapped with a filtered image in the horizontal direction of the image along the [110] and [100] zone axis, respectively. The strain is indicated by color coded in \%. In all panels the dashed lines indicate the position of the stacking fault related to the dislocation.}
    \label{fig:BaO}
\end{figure*}

The fabrication of an energetically stable BBO film is required before the Y-Bi-O system can be deposited on top. In literature, buffer layers are also used to fabricate a crystalline BBO film on a STO substrate, since the lattice mismatch between the perovskite BBO film and the STO substrate is about 12\%. A 2 nm thick BaO buffer layer was used by Makita \textit{et al.}~[17], causing the BBO film to be mainly oriented in the [100] direction instead of in both the [100] and [110] directions. Lee \textit{et al.} [13] even uses two materials, BaCeO$_{3}$ (4.4 \r{A}) and BaZrO$_{3}$ (4.19 \r{A}), to function as one buffer layer to obtain a high-quality epitaxial BBO film.

Here, a high quality BBO buffer layer is grown on top of the STO substrate without the use of an extra buffer layer. The RHEED patterns along the [100] and [110] directions of the substrate are shown in figure \ref{fig:BBOTEM}(a) for the top and bottom image, respectively. The crystalline quality and interface structure of the BBO films have been investigated using an aberration-corrected scanning transmission electron microscope (STEM), a Titan 80-300 which operated at 300kV with a convergence angle of 45-97 mrad at the collection angle for Hagh Angle Annular Dark Field  (HAADF) imaging. The specimen was prepared from a sample consisting of the STO substrate, a BBO buffer layer with a thickness of approximately 82$\pm$2~nm and a 26$\pm$2~nm thick film of the Y-Bi-O system. The Y-Bi-O film is metastable, therefore a FIB lamella was prepared in a vacuum transfer box and transferred inside a glove box to a vacuum transfer holder as described elsewhere [23-25]. An HAADF-STEM image of the interface between the STO substrate and the BBO film is shown in figure \ref{fig:BBOTEM}(b), the structure relaxes within a four unit cell interfacial layer. The HAADF-STEM image is a result of the alignment of 20 fast-acquired images of 4096x4096 pixels, aligned in the same manner as discussed by Gauquelin \textit{et al.} [26] to remove correct scanning artefacts, noise and sample drift. At the interface, every ninth unit cell shows a dislocation, which is consistent with the lattice mismatch of 12\% between the BBO and STO. This will be discussed into detail in a subsequent section. The film is stoichiometric, as observed by the \textit{in-situ} XPS. The element ratio for Ba:Bi:O is determined to be 21:25:53$\pm$3\%, respectively. The XPS data is shown in figure S1(b) of the supporting information.

\begin{figure*}[htb]%
\includegraphics*[width=\textwidth]{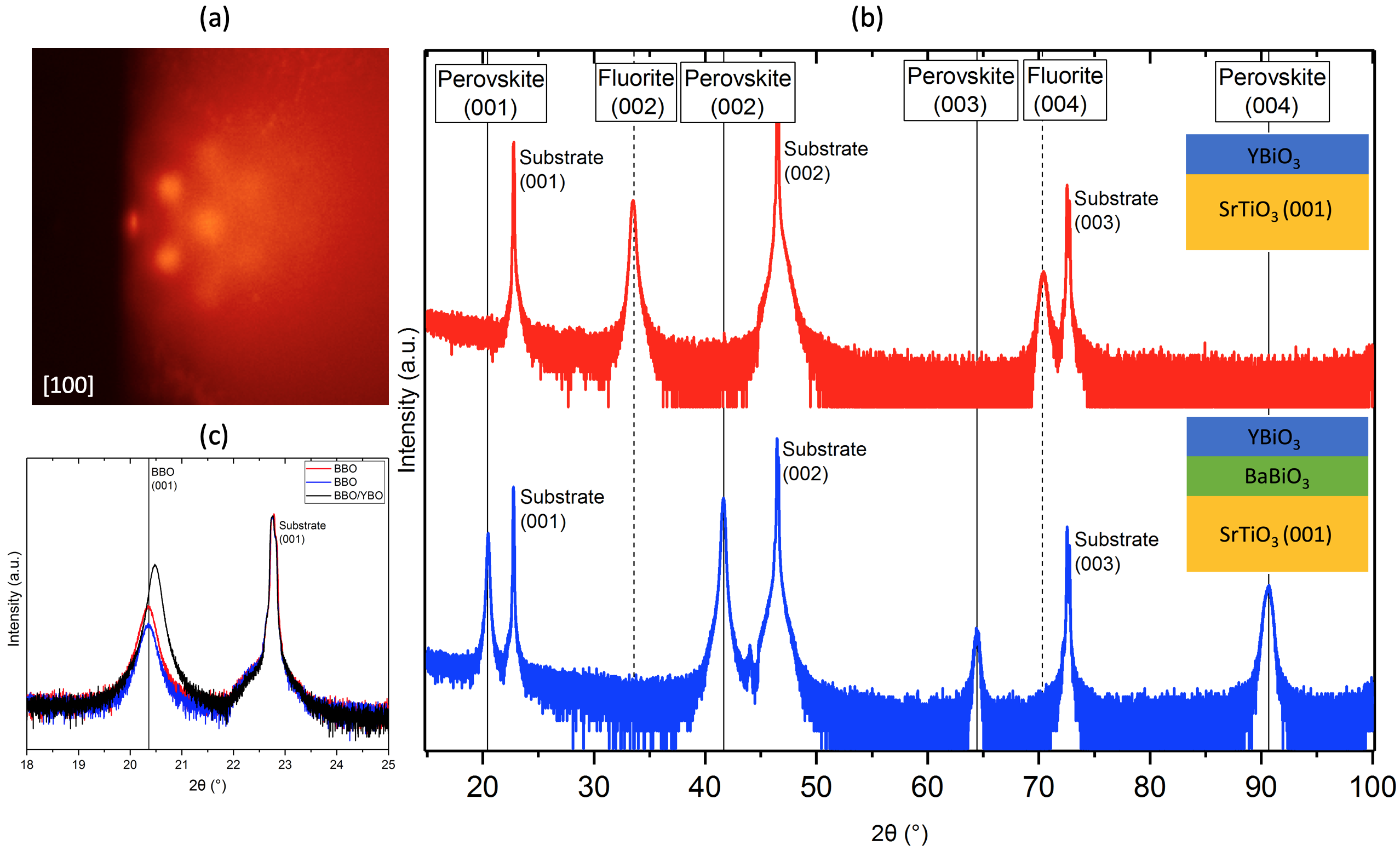}
\caption{(a) RHEED pattern of an YBO film grown on top of a BBO buffer layer, indicating a perovskite structure. (b) 2$\theta$-$\omega$ scans of an YBO film deposited directly on the STO substrate (the red curve) and an YBO film on top of a BBO buffer layer (the blue curve). The vertical dashed lines show the expected positions of a fluorite structure with a lattice constant of 5.4~\r{A} and the solid vertical lines indicate the positions for a perovskite structure with a lattice constant of 4.35~\r{A}. The schematic figures on the right, show the thin films corresponding to the scan. In (c) the intensities of the (001) diffraction peaks of two single BBO films are compared with the intensity of a BBO/YBO bilayer of which the BBO film has the same thickness as the single layers.}
    \label{fig:YBOXRD}
\end{figure*}

Anti-phase boundaries with a step of half a unit cell are observed in the BBO film. In figure \ref{fig:BBOTEM}(c) the step is indicated by the two black horizontal lines. Both black lines follow a row of bitmuth atoms, which are brighter since they are heavier than the bismuth atoms. Where the two lines are supposed to connect, a shift of half a unit cell is observed. The blue and orange dots in figures \ref{fig:BBOTEM}(b) and (c) represent the barium and bismuth atoms, respectively. The anti-phase boundary is caused by the reconstruction layer at the STO/BBO interface, BBO no longer has a single terminated nucleation side. A similar phenomena in epitaxial BBO films is discussed in the recent article by Zapf \textit{et al.} [27]. 

The lattice mismatch of BBO on STO is accommodated by dislocations every ninth unit cell and quickly resolved by two AO layers in between the two perovskite structures, whereas the first AO layer follows the stacking of the substrate as schematically visualized in figures \ref{fig:BaO}(a) and (d). The strategy of the structure to accommodate for the lattice mismatch is by introducing a second AO layer to decouple the substrate from the film. Further, we can remark notice that the dislocation core is situated exactly in this second AO layer. The intermediate layers are described as AO layers, because due to strain, the chemical composition cannot be directly assessed from the HAADF contrast and these layers might be intermixing of Ba and Sr atoms. 

\begin{figure*}[htb]%
\sidecaption
\includegraphics*[width=0.79\textwidth]{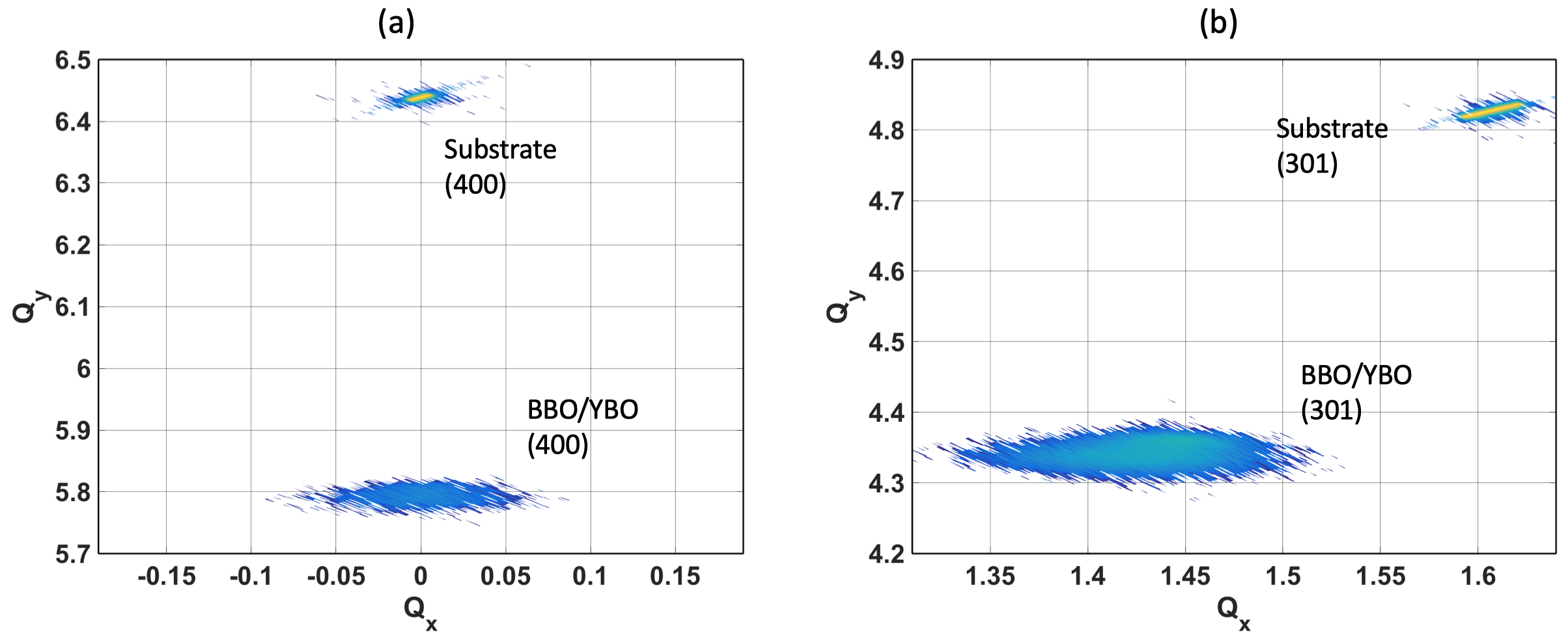}
\caption{(a) and (b) are RSMs scanned with a 2$\theta$ range of 40$^{\circ}$ and an $\omega$ range of 20$^{\circ}$, in the (400) and (301) direction of the substrate, respectively. The smaller and more intense peaks are caused by the substrate and the broader, less intense ones originate from both the BBO and YBO films.}
    \label{fig:YBOPerov}
\end{figure*}
In figure \ref{fig:BaO}, schematic visualizations and HRSTEM images of the STO/BBO interface in the [110] and [100] directions are shown. The dotted lines in both (a) and (b), indicate where the cation site of the substrate and bottommost AO layer is in between two cation site of the top AO layer. These positions form the cores of the dislocations. In (d) and (e) the dotted lines also indicate the cores of the dislocations. For easier visualization of the dislocation structure at the interface, geometric phase analysis has been used to determine strain in the in-plane direction ($\varepsilon_{xx}$). These images are overlapped with filtered images using the [1$\overline{1}$0] reflection (for the image taken along the [110] zone axis) and [010] reflection (for the image taken along the [100] zone axis) as, respectively, shown in (c) and (f). 

The two AO layers form a rocksalt structure. As in Ruddlesden-Popper compounds a perovskite ABO$_{3}$ is alternated with a rocksalt AO layer along the out-of-plane axis. Both experimentally and theoretically it is shown that by rearranging the order of AO and BO$_{2}$ layers, the energy of the film can be lowered [28, 29]. For the BBO buffer layer, the double AO layer accommodates the large lattice mismatch between the STO substrate and BBO film. The large strain is relieved by dislocations in the second AO plane which occur every ninth unit cell. The double AO is made possible by the volatile character of the bismuth atom.

When an Y-Bi-O film is deposited on top of the BBO buffer layer, a perovskite transmission pattern is observed in the RHEED image as shown in figure \ref{fig:YBOXRD}(a), comparable to the pattern in figure \ref{fig:BBOTEM}(a). As mentioned before, the Y-Bi-O film has metastable character. Therefore, it was not possible to image the structure of the Y-Bi-O film with the HRSTEM. In figure \ref{fig:BBOTEM}(c), the Y-Bi-O film is partly shown and only a distorted structure is obtained. For this reason, no simulations were performed for the Y-Bi-O layer. The 2$\theta$-$\omega$ scans of the YBO film deposited directly on a STO substrate and on top of the BBO buffer layer are both shown in figure \ref{fig:YBOXRD}(b) by the red and blue curves, respectively. The expected diffraction peak positions of the fluorite and perovskite structures are indicated by the vertical dashed and solid lines, respectively. The YBO film on the substrate solely shows diffraction peaks at the fluorite (002) and (004) positions and the three substrate diffraction peaks. The (001) and (003) diffraction peaks are absent, since they are forbidden by the structure factor of a fluorite structure. The out-of-plane lattice constant is determined to be 5.36$\pm$0.05 \r{A}.

Now focussing on the blue curve in figure \ref{fig:YBOXRD}(b), where the Y-Bi-O system is deposited on top of the BBO buffer layer, the diffraction peaks of the fluorite structure are no longer observed. Besides the substrate diffraction peaks, four diffraction peaks of the perovskite structure are appearing. The structure factor of a perovskite phase does not allow (001) and (003) either, similar to the fluorite structure. However, they are observed in this case since the atomic radius of the barium/yttrium atom is larger than the bismuth radius. In figure \ref{fig:YBOXRD}(c) the intensities of the (001) diffraction peaks for two single BBO films (the blue and red curves) and a BBO/YBO bilayer (the black curve) are compared. The BBO film of the bilayer has the same thickness as the two single BBO films. Even though the intensity of the substrate diffraction peak is the same in all three cases, the intensity of (001) diffraction peak for the BBO/YBO bilayer is significantly higher than for the two single BBO films. This clearly shows that both BBO and Y-Bi-O contribute to the perovskite diffraction peaks observed by XRD. 

In figure S2 of the supporting information, the 2$\theta$-$\omega$ scans are plotted in a stacked manner to more easily distinguish the diffraction peak details. The observed shift to higher 2$\theta$ values for the BBO/YBO compared to the single BBO films, indicates a smaller c-axis which is what one would expect for a strained film with a smaller A-site cation. This shift to higher 2$\theta$ values due to the formation of the perovskite Y-Bi-O film implies that the line shape of the diffraction peaks should be asymmetric. In the supporting information, additional analysis on the XRD data shows that the line shape of the (004) diffraction peak of the BBO/YBO bilayer is indeed asymmetric and that of the single BBO films is symmetric.

The 2$\theta$-$\omega$ scans confirms the perovskite structure indicated by the RHEED pattern. The out-of-plane lattice constant of the perovskite Y-Bi-O system is calculated to be 4.34$\pm$0.05~\r{A}. Further characterization with XRD showed that the in-plane orientation of the BBO and Y-Bi-O films was corresponding with the substrate, see the azimuthal scan in figure \ref{fig:YBOPhi}(b). Both the substrate and films $<$101$>$ peaks appear at the same positions, confirming all layers have a perovskite structure and a four-fold symmetry. 

Reciprocal space maps (RSMs) of the YBO film deposited on top of the buffer layer also confirm the perovskite structure. The RSM in figure \ref{fig:YBOPerov}(a) shows the (004) diffraction peaks of the substrate (top) and the YBO film (bottom), the BBO buffer layer also contributes to the intensity of the latter. In figure \ref{fig:YBOPerov}(b) the [103] direction is scanned, where the smaller diffraction peak is caused by the STO substrate and the bigger one by both the BBO and Y-Bi-O films. Ideally, a peak separation is expected to be observed. However, the resolution of the RSMs is too low to observe this separation. When a straight line is drawn from the substrate peak to the origin the line also crosses the centre of the film peak, indicating a fully relaxed film. This is in good agreement with the STEM analysis of the BBO film. From the RSMs the in- and out-of-plane lattice constants can be calculated, which are determined to be 4.34$\pm$0.1 \r{A} and 4.35$\pm$0.1 \r{A}, respectively. Since the penetration depth of the XPS is less than the thickness of the Y-Bi-O film deposited on top of the BBO buffer layer, the element ratio of the perovskite Y-Bi-O film can be determined. The result of the XPS scan is presented in figure~S1(c) of the supporting information and the element ratio is determined to be 23:19:58$\pm$3\% for Y, Bi and O, respectively. Electron energy loss spectroscopy (EELS) images are shown in figure S6 of the supporting information, it is concluded that no intermixing took place at the STO/BBO and BBO/Y-Bi-O interfaces. So far, the STEM analysis of the Y-Bi-O film is inconclusive since the perovskite Y-Bi-O is a highly metastable phase which makes the specimen preparation process very challenging. 

\section*{Conclusions} \text{ }
In conclusion, it is shown that the use of a buffer layer can influence the crystal phase of the thin film deposited on top. BBO is chosen as a suitable candidate since it is stable in the perovskite phase and possesses a lattice constant comparable to the 4.4~\r{A} of the perovskite YBO that is predicted to be a topological insulator. When the YBO is deposited directly on the STO substrate, it stabilizes in the fluorite phase with an out-of-plane lattice constant of 5.36$\pm$0.05~\r{A}, which is also its energetically most favourable phase. To match the oxygen positions, the YBO film is rotated by 45$^{\circ}$ in-plane with respect to the substrate. 

By first depositing a BBO buffer layer on the substrate, it becomes more energetically favourable for the Y-Bi-O system to adopt and continue the perovskite structure of the underlying film than to crystallize in the fluorite structure. The large lattice mismatch between STO and BBO is accommodated by the formation of a rock-salt like double AO layer. The double AO layer accommodates the strain by clearly visible dislocations every ninth unit cell in the upper AO layer. After these AO layers, the BBO film grows in a relaxed manner until the surface of the film.

Furthermore, the lattice mismatch between the perovskite YBO and BBO is $\sim$1\%, therefore very little strain needs to be included in the structure to make the two match. Depositing the Y-Bi-O system on top of the buffer layer, a single oriented perovskite phase is observed with an out-of-plane lattice constant of 4.34$\pm$0.05~\r{A} and the expected four-fold symmetry. It remained inconclusive if the perovksite phase in the Y-Bi-O film had a YBiO$_{3}$ or BiYO$_{3}$ configuration. These findings pave the way towards the fabrication of quantum devices for testing the hypothesised topological insulating phase in perovskite YBiO$_{3}$.

\begin{acknowledgement}
The work at the University of Twente is financially supported by NWO through a VICI grant. N.G. and J.V. acknowledge financial support  from the GOA project "Solarpaint"of the University of Antwerp. The microscope used for this experiment has been partially financed by the Hercules Fund from the Flemish Government. L. Ding is acknowledge for his help with the GPA analysis.
\end{acknowledgement}

%
%

\clearpage

\renewcommand{\thefigure}{S\arabic{figure}}
\setcounter{figure}{0}

\appendix
\begin{center}
  \textbf{\large Supporting Information: Stabilization of the perovskite phase in the Y-Bi-O system by using a BaBiO$_{3}$ buffer layer}\\[.2cm]
  
\author{%
	Rosa Luca Bouwmeester\textsuperscript{\textsf{\bfseries 1}},
 	Kit de Hond\textsuperscript{\textsf{\bfseries 1}},
  	Nicolas Gauquelin\textsuperscript{\textsf{\bfseries 2}},
  	Jo Verbeeck\textsuperscript{\textsf{\bfseries 2}},
  	Gertjan Koster\textsuperscript{\textsf{\bfseries 1}},
  	Alexander Brinkman\textsuperscript{\textsf{\bfseries 1}}
  	}

  {\itshape ${}^1$MESA$^+$ Institute for Nanotechnology, University of Twente, The Netherlands\\
  ${}^2$Electron Microscopy for Materials Research (EMAT), University of Antwerp, Belgium}\\
\end{center}

\section*{X-ray photoelectron spectroscopy}
\begin{figure*}[h]       
    \mbox{\includegraphics[width=16.4cm]{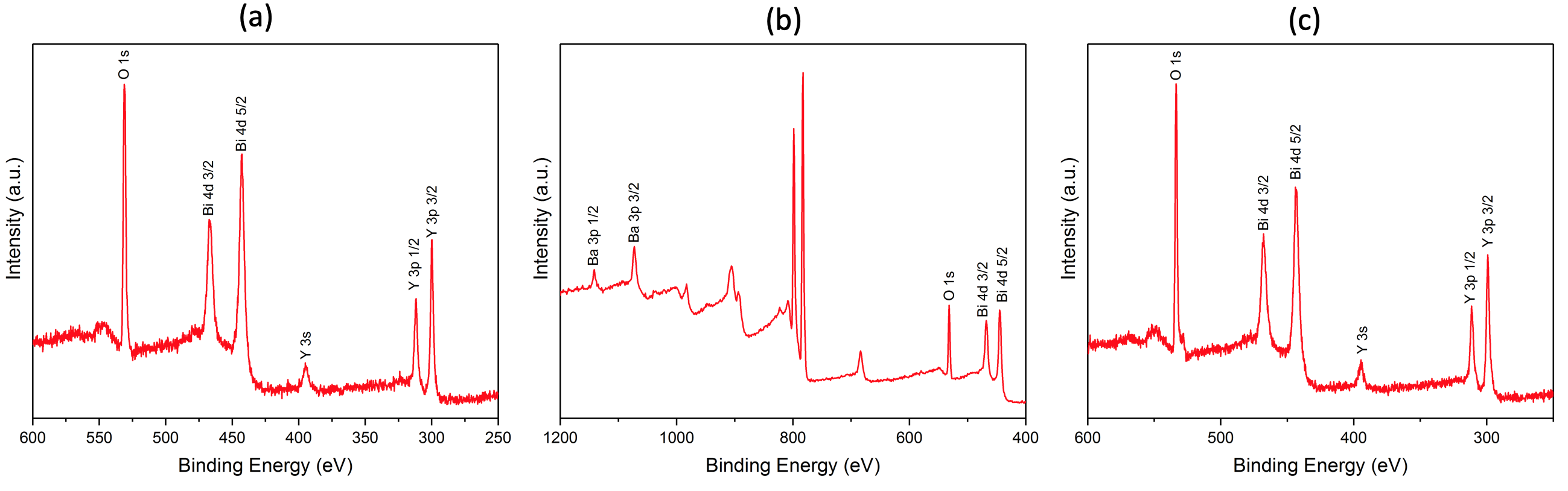}}
    \caption{The XPS results of (a) a fluorite YBiO$_{3}$ thin film, (b) a single BaBiO$_{3}$ film and (c) a BaBiO$_{3}$/YBiO$_{3}$ bilayer.}
    \label{fig:XPSSupporting}
\end{figure*}

\noindent\text{ } The Y-Bi-O system has been studied with x-ray photoelectron spectroscopy (XPS). In figure~\ref{fig:XPSSupporting} the results of a fluorite YBiO$_{3}$ thin film, a single BaBiO$_{3}$ film and a BaBiO$_{3}$/YBiO$_{3}$ bilayer are shown in (a), (b) and (c), respectively.  

\section*{X-ray diffraction}
\begin{figure*}[h]       
    \mbox{\includegraphics[width=16.4cm]{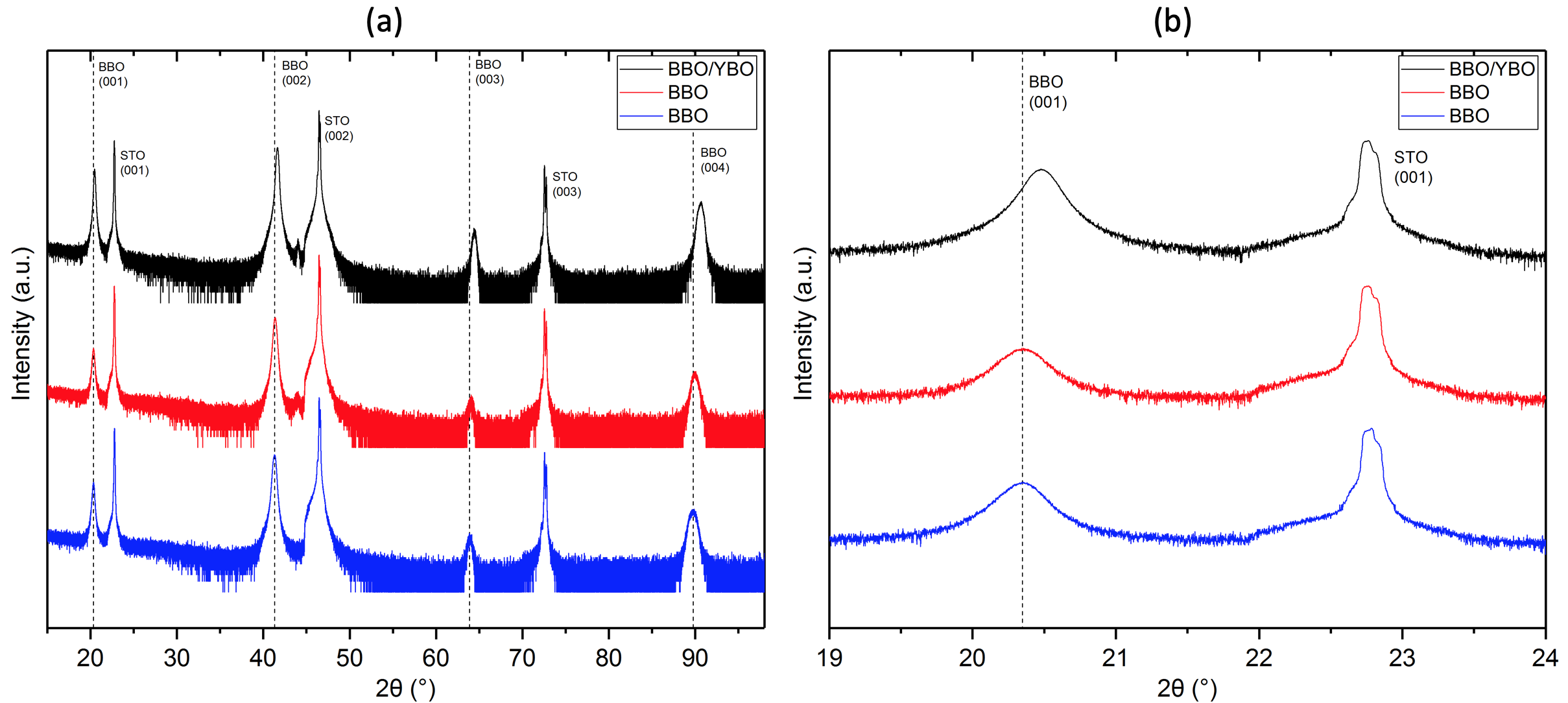}}
    \caption{The XRD 2$\theta$-$\omega$ scans of two single BaBiO$_{3}$ films (the blue and red curves) and a BaBiO$_{3}$/YBiO$_{3}$ bilayer are shown (the black curve). In (a) the full scan is shown and in (b) only the (001) diffraction peaks of the thin films and substrate are depicted.}
    \label{fig:XRDStacked}
\end{figure*}

\begin{figure*}[h]       
    \mbox{\includegraphics[width=6cm]{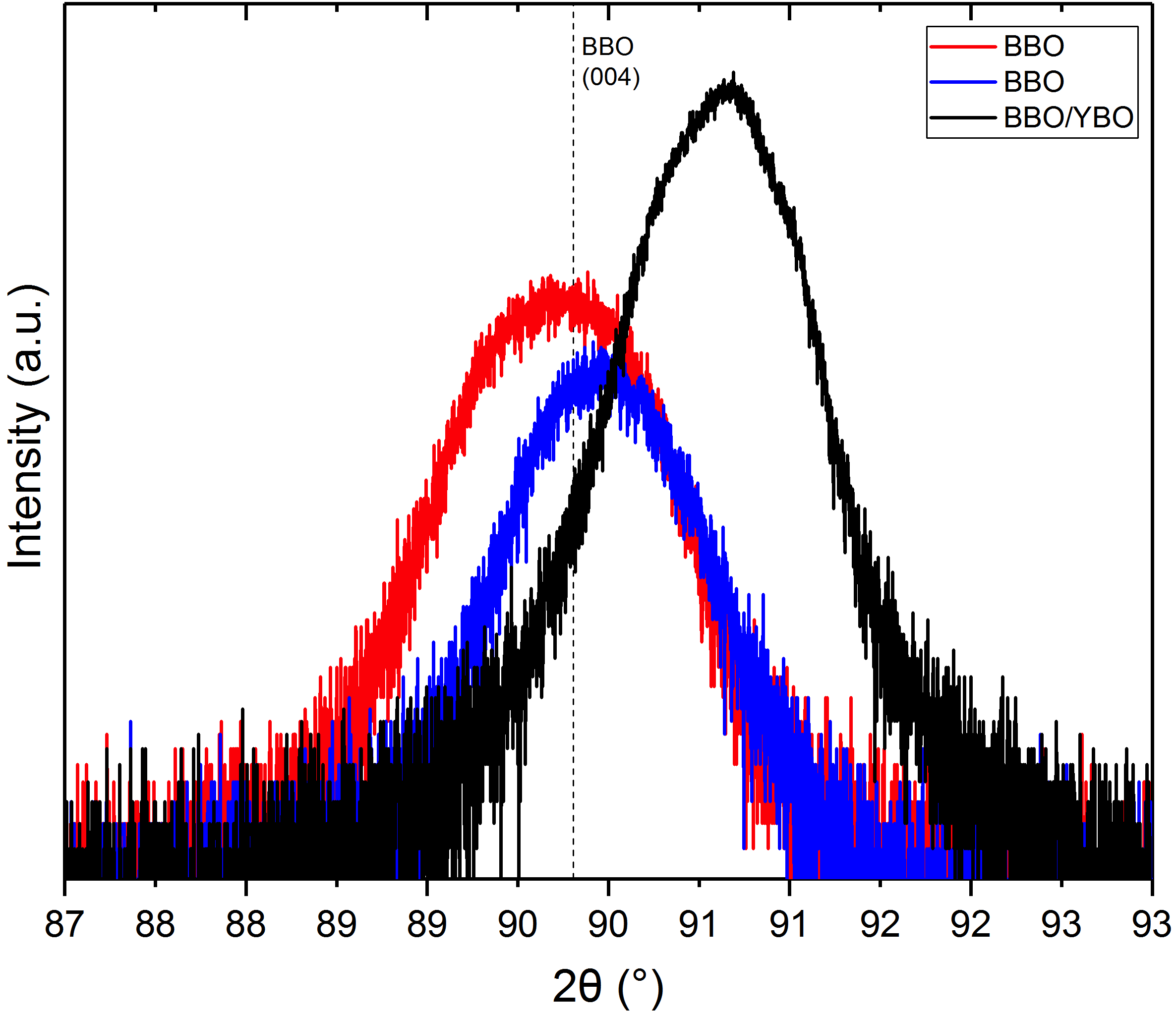}}
    \caption{The (004) diffraction peaks of XRD 2$\theta$-$\omega$ scans of two single BaBiO$_{3}$ films (the blue and red curves) and a BaBiO$_{3}$/YBiO$_{3}$ bilayer are shown (the black curve).}
    \label{fig:XRDStacked004}
\end{figure*}

\noindent\text{ } 2$\theta$-$\omega$ scans of single BaBiO$_{3}$ films and BaBiO$_{3}$/YBiO$_{3}$ bilayers are compared. In figure \ref{fig:XRDStacked} the scans of two single BaBiO$_{3}$ films and a BaBiO$_{3}$/YBiO$_{3}$ bilayer are plotted in a stacked manner, in (a) the full region is depicted and in (b) only the (001) diffraction peaks of the thin films and substrate are shown. 
\\ \\
The observed shift to higher 2$\theta$ values is due to the formation of a perovskite Y-Bi-O films, implying that the diffraction peak of the bilayer should have an asymmetric line shape. This is easier to observe for the (004) diffraction peak, since it has a higher spatial resolution than the (001) peak. In figure \ref{fig:XRDStacked004} the (004) diffraction peak of the two single BaBiO$_{3}$ films and the BaBiO$_{3}$/YBiO$_{3}$ bilayer is shown.
\\

\begin{figure*}[h]       
    \mbox{\includegraphics[width=16.4cm]{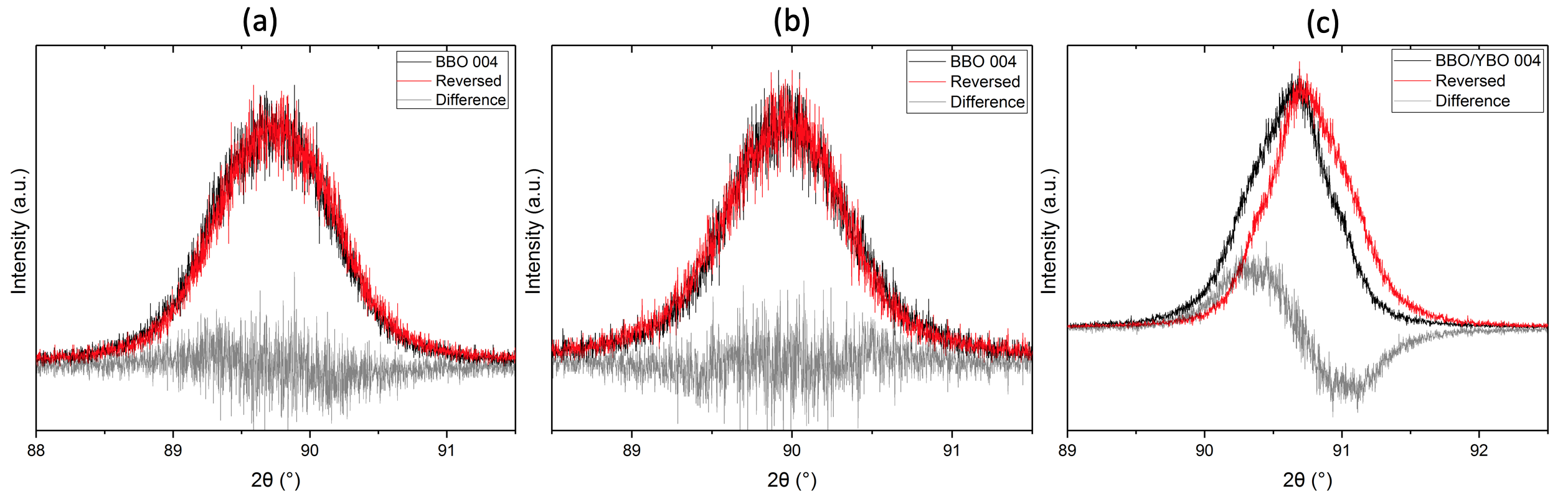}}
    \caption{The (004) diffraction peaks of XRD 2$\theta$-$\omega$ scans of two single BaBiO$_{3}$ films (a) and (b) and a BaBiO$_{3}$/YBiO$_{3}$ bilayer are shown (c). The intensities are plotted in a linear scale. The black curve shows a part of the 2$\theta$-$\omega$ scan, the red curve shows the reverse of the black curve and the gray curve is the difference between the black and red curve.}
    \label{fig:XRDSymmetry}
\end{figure*}

\begin{figure*}[h]       
    \mbox{\includegraphics[width=6cm]{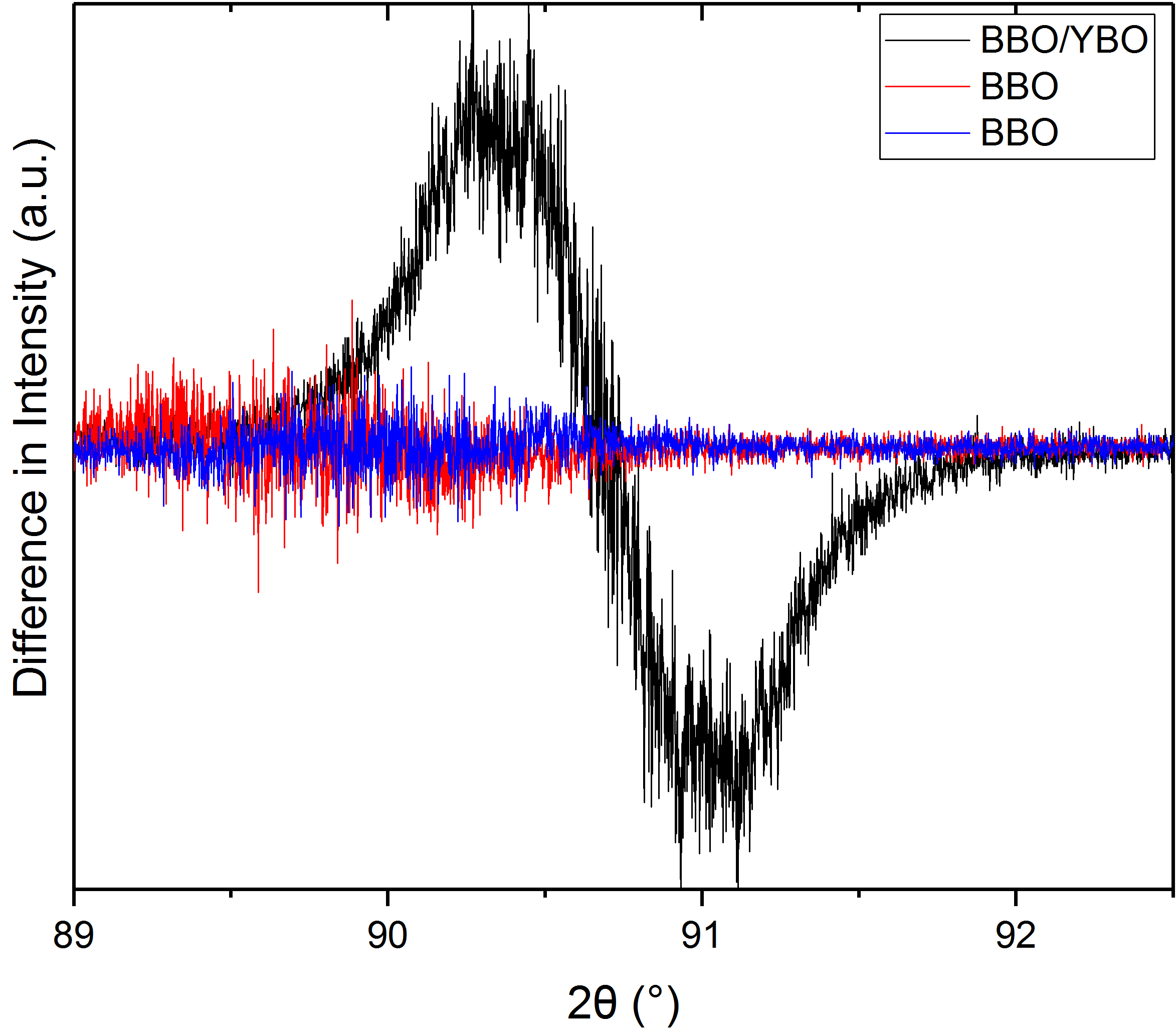}}
    \caption{The difference between the (004) diffraction peaks of the original 2$\theta$-$\omega$ scans and the reversed curve of two single BaBiO$_{3}$ films (the blue and red curves) and a BaBiO$_{3}$/YBiO$_{3}$ bilayer (the black curve).}
    \label{fig:XRDSymmetryAll}
\end{figure*}

\noindent To determine whether the line shape of the diffraction peaks is symmetric or asymmetric, an equal number of measurements points is taken on both sides of the maximum of the (004) diffraction peak. In figure \ref{fig:XRDSymmetry} this part of the scan is shown for the two single BaBiO$_{3}$ films and the bilayer in (a), (b) and (c), respectively, by the black curve. This specific part of the scan is reversed in order and plotted in red. When the black and red curve are compared, it is already clear that the line shape of the (004) diffraction peak of the single BaBiO$_{3}$ films is symmetric and that the line shape of the (004) peak of the bilayer is asymmetric. To make this more clear, the difference between the black and red curve is plotted with the gray curve. The difference between the original and reversed scans for the single BaBiO$_{3}$ films is almost equal to zero, where the difference for the bilayer deviates from zero. 
\\ \\
In figure \ref{fig:XRDSymmetryAll} the differences between the (004) diffraction peaks of the original 2$\theta$-$\omega$ scan and the reversed one are plotted, the blue and red curves show the differences of the two single BaBiO$_{3}$ films and the black curve shows the difference for the BaBiO$_{3}$/YBiO$_{3}$ bilayer. Since the difference for the singe BaBiO$_{3}$ films is close to zero, the line shape of the (004) diffraction peaks is symmetric. The line shape of the (004) diffraction peak of the BaBio$_{3}$/YBiO$_{3}$ bilayer is asymmetric, since the difference between the original and reserved 2$\theta$-$\omega$ scan is non-zero.

\section*{Electron energy loss spectroscopy}
\begin{figure*}[h]       
    \mbox{\includegraphics[width=16.4cm]{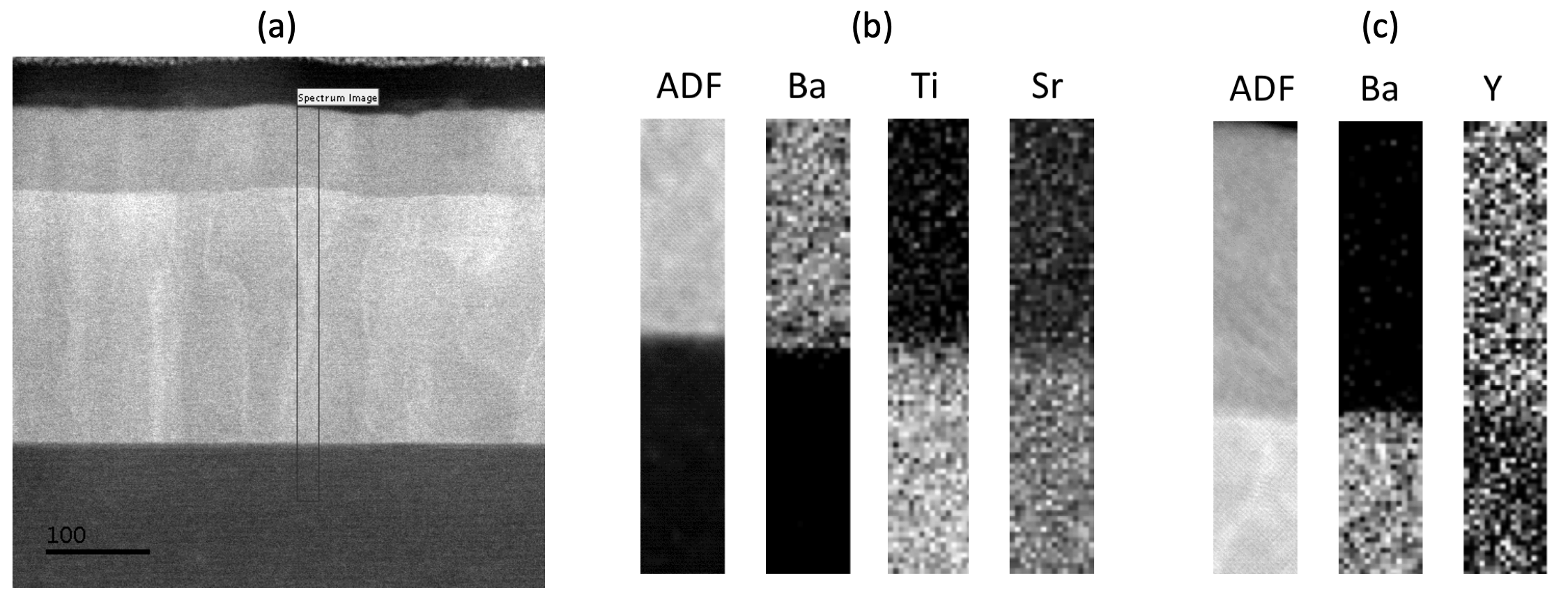}}
    \caption{EELS measurements of the specimen in annular dark-field (ADF). In (a) the measured area is indicated, where in (b) the SrTiO$_{3}$/BaBiO$_{3}$ interface is shown and in (c) the BaBiO$_{3}$/Y-Bi-O interface.}
    \label{fig:EELS}
\end{figure*}

\noindent\text{ } With the electron energy loss spectroscopy (EELS) images presented in figure \ref{fig:EELS}, the interfaces between the SrTiO$_{3}$ substrate and the BaBiO$_{3}$ film (b) and between the BaBiO$_{3}$ and Y-Bi-O films (c) are studied. The bismuth O- and M-edges are outside of the measurable range. At the SrTiO$_{3}$/BaBiO$_{3}$ interface, sharp interfaces are observed for strontium, titanium and barium atoms, all at the same height. So no intermixing took place at this interface.
\\ \\
At the BaBiO$_{3}$/Y-Bi-O interface, a sharp edge is seen for the barium atom and shows that Ba is only present below this interface. Since the yttrium atom is very small, it is harder to observe a clear boundary. However, still a change in contrast is observed at the same height as for the barium atom. So there is no interdiffusion at this interface.

\end{document}